\documentclass[twocolumn,showpacs,preprintnumbers,amsmath,amssymb]{revtex4}

\usepackage{graphicx}
\usepackage{dcolumn}
\usepackage{bm}
\usepackage{epsfig}

\begin{document}

\preprint{}

\title{High fidelity simulations of ion trajectories in miniature ion traps \\ 
using the boundary-element method}

\author{Boris Brki\'c}
\email{Boris.Brkic@liv.ac.uk}
\author{Stephen Taylor}
\author{Jason F. Ralph}
\author{Neil France}
\affiliation{Department of Electrical Engineering and Electronics, University of Liverpool, Brownlow Hill,\\ Liverpool L69 3GJ, United Kingdom}

\date{\today}

\begin{abstract}
In this paper we present numerical modeling results for endcap and linear ion traps, used for experiments at the National Physical Laboratory in the UK and Innsbruck University respectively. The secular frequencies for \textsuperscript{88}Sr\textsuperscript{+} and \textsuperscript{40}Ca\textsuperscript{+} ions were calculated from ion trajectories, simulated using boundary-element and finite-difference numerical methods. The results were compared against experimental measurements. Both numerical methods showed high accuracy with boundary-element method being more accurate. Such simulations can be useful tools for designing new traps and trap arrays. They can also be used for obtaining precise trapping parameters for desired ion control when no analytical approach is possible as well as for investigating the ion heating rates due to thermal electronic noise. 
\end{abstract} 


\maketitle

\section{\label{sec:level1}INTRODUCTION}

The manipulation of laser-cooled, trapped ions within radio-frequency (RF) traps has been widely studied for atomic optical frequency standards \cite{Rai92,Bar99}. In the last decade, ion traps have also been used for quantum information processing. This began with the first general method for the implementation of quantum-logic gates with ion traps \cite{Cir95} and hardware realization of the C-NOT gate \cite{Mon95}. Then the multiplexing trap scheme \cite{Win98} was proposed to improve the scalability, which is an important issue for quantum computing. This was followed by an architecture for a large-scale ion-trap quantum computer \cite{Kie02} and a dual linear trap array for transferring ions from one trap to another to allow sequential quantum-logic operations to be performed \cite{Row02}. 

More recently, research has turned towards the miniaturization of ion traps and investigation of different electrode geometries for microfabrication of trap arrays \cite{Mad04,Chi05}. Typically, the trap electrodes are suggested to be planar since they are much easier to fabricate than circular and hyperbolic geometries, especially at such a small sizes. These micro-trap arrays would allow the construction of more complicated devices for ion trapping at microscopic level, which should satisfy some of the main requirements for scalable quantum computing. Such miniaturization techniques can also be used for the size reduction of miniature quadrupole mass filters \cite{Tay01}. Quadrupole mass filters are mass analyzers like ion traps and they can also be constructed to form arrays.

Analytical and numerical potential modeling is crucial for ion trap design since static and ponderomotive potentials are responsible for ion oscillations within the trap. Numerical modeling of ion trap electrostatics has already been done by using different approaches, which include finite-difference method (FDM) \cite{Gul03} and finite-element method (FEM) \cite{Mad04}. Both FDM and FEM can produce inaccurate potentials and fields, particularly near the edges of electrodes where the results can be highly inaccurate. Such inaccuracy near electrode edges could have an effect on mass spectra in ion trap mass spectrometers where ions often approach electrode edges. High accuracy can also be an important factor for performance predictions in complicated electrode structures like trap arrays. Even if small inaccuracies are made in calculating suitable electrode dimensions and distances, the errors can accumulate for multi-zone trap arrays \cite{Win05}. This could later have an effect on ion transfer between different regions and ion motional heating. For these reasons, the boundary-element method (BEM) has also been considered for numerical modeling of quantum computing ion traps \cite{Brk04}. Unlike FEM and FDM, which use the whole electrode volumes to define grid points, BEM uses only the surface of electrode volumes since there is no need to consider the effects of charges below the surface. This enables faster computation and higher accuracy even with small number of electrode segments. The direct comparison of BEM, FDM and FEM has been demonstrated in Cubric et al. \cite{Cub99} for an ideal spherical analyzer and double cylinder lens using different benchmark tests. These tests included simulations with commercially available programs CPO (BEM) \cite{Cpo} and SIMION (FDM) \cite{Sim}. All the results showed that BEM had much smaller error levels than FDM and FEM for modeling potentials, fields and particle trajectories. 

In Sec. II and Sec. III, the ion secular frequencies were calculated from numerically obtained ion trajectories for both an endcap and circular linear trap. Endcap traps are used for atomic clocks and optical frequency standards at NPL \cite{Bar04}. Circular linear traps were used for quantum computing experiments at Innsbruck \cite{Nag00}. Simulation results were produced from CPO and SIMION and compared with experimentally measured secular frequencies \cite{Sin05,Nag98,Roh01}. CPO produced closer results to measurements than SIMION for both traps. The heating rate due to Johnson noise was also calculated for the Innsbruck linear trap.
 
\section{ENDCAP TRAP}

The endcap trap, proposed by Schrama et al. \cite{Sch93}, is a different geometric variant of a conventional quadrupole Paul trap and is a non-linear trap. Fig. 1 shows a model of the endcap trap used at NPL for atomic clocks and frequency standards experiments with \textsuperscript{88}Sr\textsuperscript{+} \cite{Bar04} and \textsuperscript{171}Yb\textsuperscript{+} \cite{Bly03}. A conventional quadrupole ion trap consists of one ring electrode (the hyperbolic cylinder) to trap ions in radial direction and two endcap electrodes (hyperbolic plates) for capturing ions in the axial direction. To produce an ideal quadrupole field, an RF potential is applied to the ring electrode, while the endcap electrodes are grounded. As it can be seen from Fig. 1, the NPL endcap trap has two inner endcap electrodes and two outer endcap electrodes concentric with inner ones. The outer endcaps are moved back to allow sufficient space for laser access. Equal RF voltages are applied to the inner endcaps, while small DC voltages can be applied to the outer endcaps, which are normally grounded. In effect, the inner endcaps confine ions in the axial ($z$) direction, while the outer endcaps prevent ions from escaping in the radial ($x$,$y$) direction. The equipotential lines in $zx$/$zy$ planes are shown in Fig. 2.

\begin{figure}
\begin{center}
\includegraphics[width = 6cm, height = 7cm]{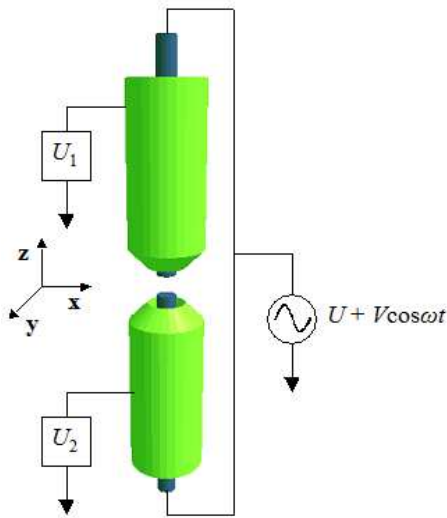}
\end{center}
\caption{(Color online) Schematic diagram of NPL endcap trap.}
\end{figure}

The mathematical theory of an endcap trap can be described through the analysis of a conventional quadrupole ion trap. Since the electric field in quadrupole ion traps is rotationally symmetric, we can represent radial axes $x$ and $y$ with $r$ (= $\sqrt{x^{2} + y^{2}}$). Thus, the trap potential at any point in a conventional quadrupole ion trap is given by:
\begin{equation}
\Phi{(r,z)} = \frac{r^{2} - 2z^{2} + 2z_{0}^{2}}{r_{0}^{2} + 2z_{0}^{2}}{(U + V\cos(\Omega{t}))},
\end{equation}
where $U$ and $V$ are the DC voltage and zero-to-peak AC amplitude applied to the ring electrode, $\Omega$ is the angular frequency equal to 2$\pi{f}$, where $f$ is the frequency of the RF field, $r_{0}$ is the smallest distance from the trap center to the ring electrode and $z_{0}$ is the smallest distance from the trap center to the endcap electrode. For an ideal quadrupole field the ratio between $r_{0}$ and $z_{0}$ is given by:
\begin{equation}
{r_{0}^2} = 2{z_{0}^2}.
\end{equation}

\begin{figure}
\begin{center}
\includegraphics[width = 6.5cm, height = 4cm]{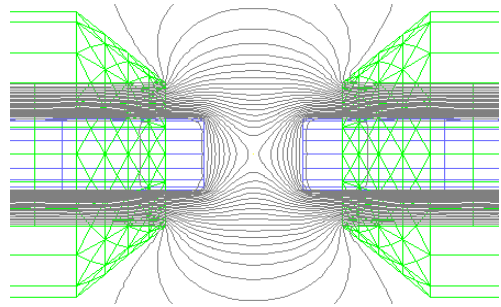}
\end{center}
\caption{(Color online) Equipotential contours in $zx$($zy$) plane for the NPL endcap trap.}
\end{figure}

The equations of motion for an ion at mass $m$ and charge $e$ are given by:
\begin{eqnarray}
\frac{{d}^{2}r}{d{t}^2} + \frac{e}{2m{z_{0}^{2}}}(U + V\cos(\Omega{t}))r = {0}, \nonumber \\
\frac{{d}^{2}z}{d{t}^2} - \frac{e}{m{z_{0}^{2}}}(U + V\cos(\Omega{t}))z = {0}.
\end{eqnarray}
Ion stability parameters $a_{u}$ and $q_{u}$ are obtained by solving the Mathieu equation:
\begin{equation}
\frac{{d}^{2}u}{d{\xi}^2} + (a_{u} - 2q_{u}\cos(2\xi))u = {0},
\end{equation}
where $u$ can be either $x$, $y$ or $z$ and $\xi = (\omega{t})/2$. The resulting expressions for $a_{u}$ and $q_{u}$ are:
\begin{eqnarray}
{a_{x,y}} = -\frac{1}{2}{a_{z}} = \frac{2{\varepsilon}eU}{m{z_{0}^2}{\Omega}^2}, \\
{q_{x,y}} = -\frac{1}{2}{q_{z}} = -\frac{{\varepsilon}eV}{m{z_{0}^2}{\Omega}^2},
\end{eqnarray}
where $\varepsilon$ is the `efficiency' of the trap. A conventional quadrupole ion trap produces an ideal quadrupole field and it has $\varepsilon$ = 1. It was shown experimentally that the NPL endcap trap has $\varepsilon$ = 0.63 \cite{Bly03}. 

\begin{table*}
\caption{\label{tab:table1}Experimental and numerical results for secular frequencies of a \textsuperscript{88}Sr\textsuperscript{+} ion, trapped within the NPL endcap trap.}
\begin{ruledtabular}
\begin{tabular}{ccccccc}
 Secular frequencies & Description & 1st set & 2nd set & 3rd set & 4th set & 5th set \\ 
 &\multicolumn{1}{c}{} &\multicolumn{1}{c}{199 V RMS} &\multicolumn{1}{c}{221 V RMS} &\multicolumn{1}{c}{245 V RMS}   
 &\multicolumn{1}{c}{274 V RMS} &\multicolumn{1}{c}{304 V RMS} \\ 
 &\multicolumn{1}{c}{} &\multicolumn{1}{c}{15.955 MHz} &\multicolumn{1}{c}{15.948 MHz} &\multicolumn{1}{c}{15.936 MHz} &\multicolumn{1}{c}{15.925 MHz} &\multicolumn{1}{c}{15.91 MHz} \\
 &\multicolumn{1}{c}{} &\multicolumn{1}{c}{2.12 V DC} &\multicolumn{1}{c}{2.55 V DC} &\multicolumn{1}{c}{3.31 V DC} &\multicolumn{1}{c}{2.39 V DC} &\multicolumn{1}{c}{2.38 V DC} \\ \hline
 $\omega_{x,y}/{2\pi}$ & {Experiment} & {1.395 MHz} & {1.590 MHz} & {1.800 MHz} & {1.980 MHz} & {2.230 MHz} \\
 $\omega_{z}/{2\pi}$ & {Experiment} & {2.985 MHz} & {3.360 MHz} & {3.795 MHz} & {4.340 MHz} & {5.070 MHz} \\
 $\omega_{x,y}/{2\pi}$ & {BEM} & {1.403 MHz} & {1.596 MHz} & {1.789 MHz} & {1.980 MHz} & {2.227 MHz} \\
 $\omega_{z}/{2\pi}$ & {BEM} & {2.939 MHz} & {3.265 MHz} & {3.767 MHz} & {4.281 MHz} & {4.960 MHz}\\
 $\omega_{x,y}/{2\pi}$\ & {FDM} & {1.441 MHz} & {1.606 MHz} & {1.791 MHz} & {1.988 MHz} & {2.213 MHz} \\
 $\omega_{z}/{2\pi}$ & {FDM} & {2.879 MHz} & {3.247 MHz} & {3.668 MHz} & {4.261 MHz} & {4.946 MHz}\\
\end{tabular}
\end{ruledtabular}
\end{table*}

An ion is stable within the ion trap if it has a stable trajectory in both radial and axial directions. Another important trapping parameter is ${\beta}_{u}$, which depends on $a_{u}$ and $q_{u}$ and 0 $< {\beta}_{u} <$ 1 must hold. In order to obtain the exact value of ${\beta}_{u}$, a continued fraction in terms of $a_{u}$ and $q_{u}$ must be used. A simpler expression for ${\beta}_{u}$ is the Dehmelt approximation given by \cite{Mar97}:
\begin{equation}
{\beta}_{u} = \left[a_{u} + (q_{u}^2 / 2)\right]^{1/2}.
\end{equation}
which is only valid for $q_{x,y} <$ 0.2 and $q_{z} <$ 0.4. Because of its trap parameters (e.g. RF voltages), the NPL endcap has larger values of  $q_{u}$ and the Dehmelt approximation cannot be used. Therefore, the fourth order approximation for ${\beta}_{u}$ should be used \cite{Hoff01}:
\begin{equation}
{\beta}_{u} = \left[a_{u} - \frac{(a_{u} - 1)q_{u}^2}{2(a_{u} - 1)^2 - q_{u}^2} - \frac{(5a_{u} + 7)q_{u}^4}{32(a_{u} - 1)^3(a_{u} - 4)}\right]^{1/2}.
\end{equation}

Ion motion in the RF field consists of secular motion (slow oscillations) and micromotion (fast oscillations). In an endcap trap, the ion experiences micromotion in all three directions. Micromotion can cause adverse effects for laser-cooled ions, such as significant second-order Doppler shifts when high-accuracy is investigated and limited confinement time in the absence of cooling. This is due to the increase of ion motional heating. In experiments, micromotion is normally minimized with the fluorescence modulation technique \cite{Ber98} and compensation electrodes, which are used to move the ion towards the trap center where the energy of micromotion is the lowest. The NPL endcap trap has two compensation electrodes orthogonal to each other and to the trap electrodes, which reduce micromotion in radial direction, while small DC voltages can be applied to outer endcap electrodes for the reduction of micromotion in the axial direction. The ion motional frequencies are normally called secular frequencies, since micromotion is very small compared to secular motion and its influence can be neglected for high frequencies. The expression for angular secular frequencies is an algebraic progression and it is given by \cite{Mar97}:
\begin{equation}
{\omega}_{u,n} = \left(n \pm \frac{\beta_{u}}{2}\right)\Omega, ~~~~~ 0 \leq n < \infty,
\end{equation} 
where $n$ represents the frequency order. The angular secular frequency at $n$ = 0 is:
\begin{equation}
{\omega}_{u} = \frac{\beta_{u}\Omega}{2},
\end{equation}
and it is called the fundamental frequency, having the lowest value of all orders and the highest power spectrum. The expressions for ${\omega}_{x,y}$ and ${\omega}_{z}$ can be obtained by using the approximation for $\beta_{u}$ from Eq. (8) and placing it into Eq. (10). The inner endcap DC voltage $U$ is always set to zero, so that $a_{u}$ = 0 for the NPL endcap trap and only $q_{u}$ is necessary for calculations of $\beta_{u}$.

To model ion traps accurately, all the trap parameters need to be explicitly specified. This includes both trap dimensions and driving voltages. The NPL endcap trap consists of two inner and two outer endcap electrodes made from Tantalum, which have an Alumina insulation spacer between them. The inner encaps have 0.5 mm diameter and a length much larger than their diameter (approx. 16 mm). The outer endcaps have 1 mm inner diameter and 2 mm outer diameter. The inner endcaps are separated from each other by 0.56 mm, which is equal to 2$z_{0}$. The outer endcaps are separated by 1 mm and angled at 45 degrees with respect to the $z$-axis, which is also called the trap axis. In the simulations, the trap was driven with different voltages and at different frequencies corresponding to a few sets of experiments performed at NPL. A \textsuperscript{88}Sr\textsuperscript{+} ion was injected at the trap center at 0.05 eV kinetic energy with equal initial velocities in all directions and allowed to oscillate for 1 ms.

Table I shows the numerical and experimental values for secular frequencies of a \textsuperscript{88}Sr\textsuperscript{+} ion for five different sets of experiments at NPL. In all the experiments a single laser-cooled ion was confined near the trap center. Each set contains different trapping parameters. These include the RF voltages and driving frequencies applied to the inner endcaps and small DC voltages applied to one outer endcap, while the other one remains grounded. The inner endcap DC voltages are equal to zero. The secular frequencies for an endcap trap cannot be obtained analytically like for a conventional quadrupole ion trap using Eq. (9). This is because an endcap trap does not have an ideal quadrupole field and $\varepsilon$ = 1. Thus, the value of its efficiency can only be estimated experimentally or numerically. The secular frequencies were calculated from ion trajectories produced by CPO (BEM) and SIMION (FDM), and compared with exact experimental measurements taken from \cite{Sin05}. The measurement technique used at NPL is described in \cite{Sin01}, where radial and axial secular frequencies can be seen from the experimental sideband spectrum.

Before modeling the NPL endcap trap, BEM and FDM simulations were performed on a conventional quadrupole ion trap to compare their results against theory (see Eq. (1)). These simulations showed that BEM was in average 1\% more accurate than FDM for basic potential and field calculations near the trap center. Table I also shows that on average BEM produced closer secular frequencies to the measurements for the NPL endcap trap than the ones FDM generated. Obviously this is a basic comparison and it would require more experimental sets and simulations to provide a definitive picture. It can be seen from the table that the radial frequency is smaller than the axial frequency, which is always the case for an endcap trap since the RF field is stronger in the axial direction due to the RF driven inner endcaps. 

From the secular frequencies, the DC component of the electric quadrupole field gradient can be calculated, which can be used to determine the quadrupole moment of an ion \cite{Bar04}. This can be useful when designing traps for atomic clocks and frequency standards with different ions. 

\section{LINEAR TRAP}

\begin{figure}
\begin{center}
\includegraphics[width = 8.6cm,height = 5cm]{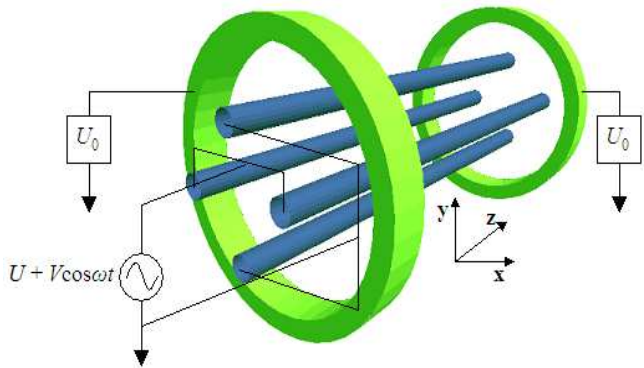}
\end{center}
\caption{(Color online) Schematic diagram of Innsbruck linear trap with circular electrodes.}
\end{figure}

A conventional linear Paul trap consists of a hyperbolic-rod quadrupole mass filter with prefilter and postfilter of the same shapes, which represent endcap electrodes. The main filter traps ions radially with RF voltages, while the prefilter and postfilter confine them axially with DC voltages. Because of the difficulty in the manufacture of hyperbolic electrodes, other geometric shapes are used for linear traps, such as circular-rod \cite{Nag00}, blade-shaped \cite{Kal03}, rectilinear \cite{Ouy04} and planar \cite{Chi05}, which have all found applications in different areas. 

\begin{figure}
\begin{center}
\includegraphics[width = 7cm, height = 4.5cm]{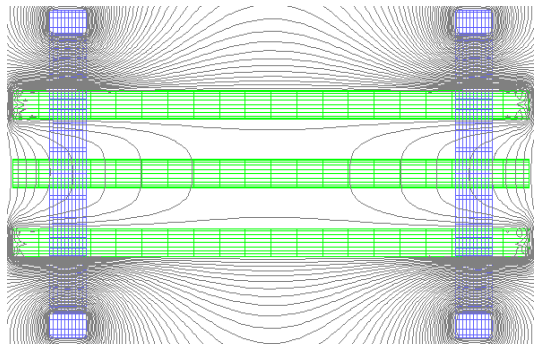}
\end{center}
\caption{(Color online) Equipotential contours in $zx$/$zy$ plane for the Innsbruck circular-rod linear trap.}
\end{figure}

This section will investigate only linear traps that are used for quantum computation. Fig. 3 shows the circular-rod linear trap, which represents the `old' trap model used for quantum computing with \textsuperscript{40}Ca\textsuperscript{+} ions at Innsbruck University. It has four circular rods with two ring-shaped electrodes placed around the ends of the quadrupole rods. Ions are radially trapped with RF voltages applied to the quadrupole rods, while axially they are confined by DC voltages applied to the endcap rings. In linear traps used for quantum computing one pair of the two diagonally opposing RF electrodes is grounded, while in linear trap mass spectrometers, the two pairs of diagonal rods have equal voltage values, but opposite in sign. The ratio between the rod thickness and diagonal rod distance is lower for quantum computing traps than for mass spectrometry traps. This is because quantum computing traps do not require ion filtering and they need larger space between the electrodes to allow access for laser beams. Also there is no DC voltage applied to the RF electrodes, since again ions do not need to be filtered, but only to remain trapped.

Linear trap theory is based on quadrupole mass filter theory. The electric fields inside a linear trap have components along the trap axis ($z$ direction) and radial axes ($x$ and $y$ directions). Fig. 4 shows the equipotential lines in $zx$/$zy$ planes for the Innsbruck linear trap. The trap potential at any point in the radial direction is given by:
\begin{equation}
\Phi{(x,y)} = \frac{x^{2} - y^{2}}{r_{0}^{2}}{(U + V\cos(\Omega{t}))},
\end{equation}
where $U$ and $V$ are the DC voltage and zero-to-peak AC amplitude applied to the quadrupole rods, $\Omega$ is the angular frequency equal to 2$\pi{f}$, where $f$ is the frequency of the applied RF field, and $r_{0}$ is the smallest distance from the quadrupole center to the surface of the electrodes. Therefore, the equations of motion for an ion at mass $m$ and charge $e$ can be derived: 
\begin{eqnarray}
\frac{{d}^{2}x}{d{t}^2} + \frac{2e}{m{r_{0}^{2}}}(U + V\cos(\Omega{t}))x = {0}, \nonumber \\
\frac{{d}^{2}y}{d{t}^2} - \frac{2e}{m{r_{0}^{2}}}(U + V\cos(\Omega{t}))y = {0}, \nonumber \\
\frac{{d}^{2}z}{d{t}^2} = {0}.
\end{eqnarray}
Ion stability parameters $a_{u}$ and $q_{u}$ are obtained by solving the Mathieu equation given in Eq. (4). The resulting expressions for $a_{u}$ and $q_{u}$ are:
\begin{eqnarray}
{a_{x,y}} = -\frac{1}{2}{a_{z}} = -\frac{4e{\kappa}U_{0}}{m{\Omega}^2{z_{0}}^2}, \\
{q_{x}} = -{q_{y}} = \frac{4eV}{m{\Omega}^2{r_{0}}^2}, ~~~ q_{z} = 0,
\end{eqnarray}
where $\kappa$ is the `geometric factor' \cite{Ber98} which must be estimated, $U_{0}$ is the DC voltage applied to both endcap rings and $z_{0}$ is the smallest distance from the trap center to the endcap electrode. In practice the geometric factor describes the intensity of the static DC field in the axial direction. Its value depends on the change of trap dimensions and the static DC voltages applied to the endcap electrodes. 

In theory, an RF field in linear traps has no components in the axial direction. However, when a whole 3D structure is modeled numerically, a small presence of RF field can be seen in the axial direction. This means that micromotion also exists in the axial direction, but it is much smaller than in radial. The expressions for radial and secular frequencies can be obtained by using the approximation for $\beta_{u}$ from Eq. (7) and substituting it into Eq. (10). Since $|a_{x,y}| << |q_{x,y}|$ for the Innsbruck linear trap, the formula for the radial angular frequency can be approximated by:
\begin{equation}
{{\omega}_{x,y}} = \frac{2eV}{\sqrt{2}m{\Omega}{r_{0}}^2}.
\end{equation}
Since $q_{z}$ = 0, the expression for the axial angular frequency is given by:
\begin{equation}
{{\omega}_{z}} = \sqrt{\frac{2e{\kappa}U_{0}}{m{z_{0}}^2}}.
\end{equation}

\begin{figure}
\begin{center}
\includegraphics[width = 8.6cm, height = 7.5cm]{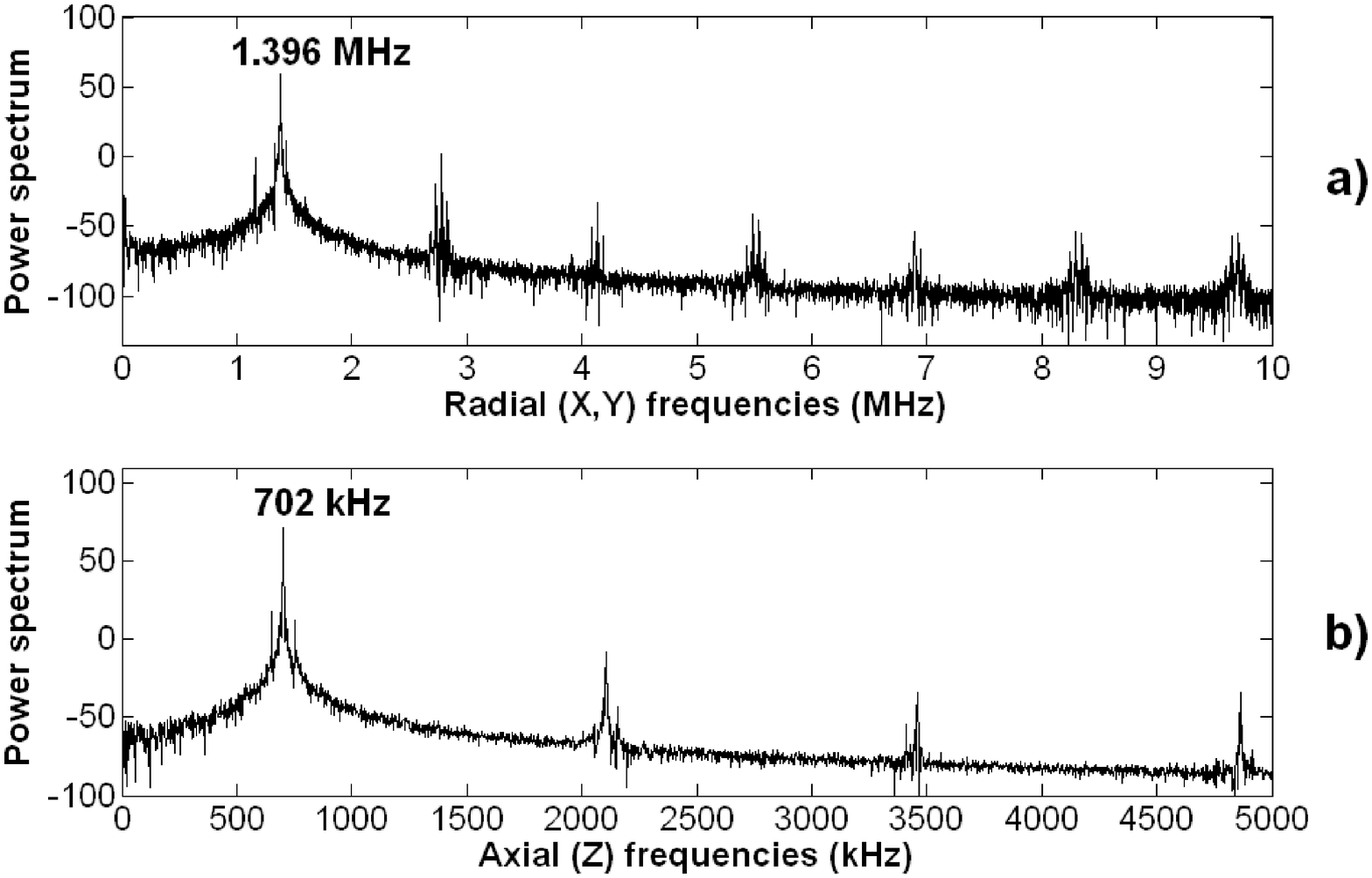}
\end{center}
\caption{BEM generated power spectra for secular frequencies of a \textsuperscript{40}Ca\textsuperscript{+} ion, trapped within the Innsbruck linear trap. The trap driven with 1000 V AC peak at 18 MHz.}
\end{figure}

The Innsbruck linear trap consists of four stainless steel cylindrical electrodes, forming a quadrupole, and two ring-shaped electrodes at the ends. The electrodes are isolated by MACOR spacers. The quadrupole rods have 0.6 mm diameters with their diagonal separation (2$r_{0}$) equal to 2.4 mm. The endcap rings are 6 mm in diameter and the distance between them (2$z_{0}$) is 10 mm. In the simulation, the trap was driven with 1000 V peak at 18 MHz applied to one pair of RF electrodes with other pair grounded and 2000 V applied to the endcap DC rings. A \textsuperscript{40}Ca\textsuperscript{+} ion was injected at the trap center at 1 eV kinetic energy with equal initial velocities in all directions and allowed to oscillate for 1 ms. 

As for an endcap trap, the secular frequencies for a linear trap cannot be obtained analytically because of the geometric factor involved. Fig. 5 shows the plots of \textsuperscript{40}Ca\textsuperscript{+} ion motional frequencies generated by BEM for given Innsbruck linear trap parameters. The power spectrum plot for axial frequencies has much less noise than the radial plot. This is because of very small presence of an RF field in axial direction. The experimental measurements gave $\omega_{x,y}/{2\pi}$ = 1.400 MHz and $\omega_{z}/{2\pi}$ = 700 kHz, which can be clearly seen from the full sideband spectrum given in \cite{Nag00,Roh01}. BEM produced $\omega_{x,y}/{2\pi}$ = 1.396 MHz and $\omega_{z}/{2\pi}$ = 702 kHz, while FDM produced $\omega_{x,y}/{2\pi}$ = 1.507 MHz and $\omega_{z}/{2\pi}$ = 696 kHz. The BEM results are again closer to the measurements than FDM results, especially for the radial frequency where the BEM showed significantly higher accuracy. It can be seen that the radial frequency is larger than the axial, which is normally the case in linear traps. 
  
The numerical prediction of the secular frequencies can be used for estimating the heating rates when designing new traps. Ion motional heating is an important issue for quantum computation. It can lead to decoherence of superposition states and increase the ion separation time in multiplexed traps, which would limit the speed of quantum-logic operations \cite{Win98,Row02}. The main causes of heating of ion motional modes include Johnson noise that can happen due to the resistance of the trap electrodes or external circuits, and fluctuating patch potentials, which are influenced by the noise from microscopic electrode regions. The heating rate due to resistance of trap electrodes is given by \cite{Tur00}:
\begin{equation}
\hbar{\omega}_{u}{\dot{\overline n}}_{u} = \frac{{e}^2kTR}{m{z}^2},
\end{equation}
where ${\overline n}_{u}$ is the average vibrational quantum number of an ion for a given direction, $T$ is the trap operation temperature (usually $T$ = 300 K), $z$ is the distance of the ion from the conductive electrode and $R$ is the electrode resistance, which can be calculated using resistivity of electrode material ($\rho$ = 7.5$\times{10}^{-6}$ $\Omega$cm for stainless steel). Thus, the Innsbruck trap heating rate for radial modes due to Johnson noise is approximately 1 quantum/670 ms (using the BEM generated secular frequency). This is obviously an underestimate for the heating rate because of the other influential factors previously mentioned. Ion motional heating can especially be increased due to fluctuating potentials on electrode surfaces. However, it has been demonstrated that smooth and pure electrode surfaces can significantly minimize the heating rate \cite{Row02}. If electrodes were made smooth enough, then the calculation of Johnson noise heating (inevitably present) will give a better estimate for the actual ion heating.

\section{CONCLUSIONS}

This paper has shown simulation results of ion secular frequencies for two different ion traps using the boundary-element method and finite-difference method. The results were compared with experimental measurements in each case and the boundary-element method proved to be more accurate than the finite-difference. We suggest that the boundary-element method should be used for accurate modeling of ion-trap electrostatics and particle trajectories. This method should especially be useful when used in the design of miniature trap arrays that could potentially be used for scalable quantum computers. Such high fidelity simulations would help to prevent design errors, which could accumulate for complicated multi-zone linear traps. Ion motional heating is also of interest and the heating rate due to Johnson noise can be calculated from numerically obtained secular frequencies. The relative permittivity of the insulating materials can be included in future simulations to check whether it has effects on ion secular frequencies and motional heating. 

For the purpose of quantum computation, a logical next step could be the numerical (using space-charge support) and analytical modeling of two or more ions oscillating simultaneously in a trap. Such modeling would include all the mutual interactions and quantum effects that ions experience during their oscillations enabling the prediction of superposition of motional states. This could lead to an ion-trap C-NOT gate simulator, which might be further expanded to simulate quantum algorithms.


\end{document}